\documentclass{article}

\usepackage{graphicx}
\usepackage{amsmath,amssymb}
\usepackage{color}

\begin{document}


\title{\vspace{-2.5cm} \bf \large Non-sinusoidal current and current reversals in a gating ratchet}

\author{Luis Dinis$^{1,2}$ and Niurka R. Quintero$^{3,4}$}
\date{}
\maketitle
\vspace{-1cm}
\begin{center}
{\em $^{1}$ Departamento de F\'{\i}sica At\'{o}mica, Molecular y Nuclear,\\ Universidad Complutense de Madrid, 28040 Madrid, Spain.\\
$^{2}$ GISC -- Grupo Interdisciplinar de Sistemas Complejos, Madrid, Spain\\
$^{3}$ Instituto de Matem\'aticas de la
Universidad de Sevilla (IMUS)\\
$^{4}$
Departamento de F\'{\i}sica Aplicada I, E.P.S., Universidad de Sevilla, 
Calle Virgen de \'Africa 7, 41011 Sevilla, Spain\\
}
\date{\today}
\end{center}
\begin{center}
\parbox{0.85\linewidth}{In this work, the ratchet dynamics of Brownian particles driven by an external sinusoidal ({\em harmonic}) force is investigated. The gating ratchet effect is observed  when another harmonic is used to modulate the 
spatially symmetric potential in which the particles move. 
For small amplitudes of the harmonics, it is shown that the {\em current} (average velocity) of particles exhibits a sinusoidal
shape as a function of a precise combination of the 
phases of both harmonics. By increasing the amplitudes of the harmonics
beyond the small-limit regime, departures from the sinusoidal
behavior are observed and current reversals can
also be induced. 
These current reversals persist even for the overdamped dynamics of the particles. 
\\
\\
{\sc pacs} numbers: {05.40.-a, 05.45.-a, 05.60.-k}
}
\end{center}

\section{Introduction} \label{intro}

The transport of particles or solitons under zero-average forces (i.e., ratchet transport) 
has been extensively investigated in the last two decades \cite{ajdari:1994,hanggi:1996,astumian:1997,reimann:2002a,hanggi:2009}. 
This phenomenon has been predicted and explained in different fields of physics, ranging from nano-devices to 
molecular motors \cite{linke:2002,reimann:2002a,hanggi:2009}. 
Moreover, it has also been observed in experiments and simulations with  nonlinear systems, where spatio-temporal symmetries  have been properly broken \cite{lee:1999,linke:1999,villegas:2003,salerno:2002,julicher:1997,ooi:2007}. 
In particular, the ratchet models were used: to elucidate the working principles of molecular motors; to design  molecular motors \cite{kay:2007}; and to explain the unidirectional motion of fluxons in Josephson junctions \cite{ustinov:2004,ooi:2007}, the transport of cold atoms in optical lattices  \cite{schiavoni:2003}, and the vortices in superconductors \cite{villegas:2003,dinis:2007}.

The ratchet transport is described by means of the {\em current} (average velocity) \cite{hanggi:1996,hanggi_brownian_1998,reimann:2002a,hanggi:2009},  
\begin{equation}
v=\lim_{t\to\infty}\frac{\langle x(t)\rangle-x_0}{t-t_0},
\label{eq:v}
\end{equation}
where $x(t)$ is the position of particles, or the center of mass of solitons at time $t$, $\langle\cdot\rangle$ represents an ensemble average over all
trajectories satisfying the same initial condition, and $x(t_0)=x_0$.


Two possible underlying mechanisms of rocking ratchets  are {\em harmonic mixing} and {\em gating}. 
The current of particles (atoms or solitons) in \emph{harmonic mixing} is generally induced by an additive bi-harmonic, T periodic,  
driving force  $f(t)=f_{1}(t)+f_{2}(t)$, with 
\begin{equation}
  f_{1}(t) = \epsilon_{1} \cos(q_1 \omega t), \qquad f_{2}(t)=\epsilon_{2} \cos(q_2 \omega t +\phi),
\label{eq:drive}
\end{equation}
where $\epsilon_1$ and $\epsilon_2$ are the amplitudes of the harmonics, $\phi$ is the relative phase between the two harmonics, 
$(q_1,q_2)\in \mathbb{N}^2$, $\text{gcd}(q_1,q_2)=1$ and $T=2 \pi/\omega$. On the other hand, in \emph{ gating ratchets}, particles experience a  symmetric potential with the amplitude modulated by means of $f_1(t)$. A time-symmetric harmonic force $f_2(t)$ is also applied.

The time-shift invariance of the current, 
\begin{equation}\label{eq:timeshift}
v[f_1(t+\tau),f_2(t+\tau)]=v[f_1(t),f_2(t)],
\end{equation} 
$\forall \tau$, together with the  symmetry
\begin{equation}\label{eq:fr1}
v[-f_1(t),-f_2(t)]=-v[f_1(t),f_2(t)],
\end{equation}
or  
\begin{equation}\label{eq:fr2}
v[f_1(t),-f_2(t)]=-v[f_1(t),f_2(t)],
\end{equation}
fix the necessary conditions on $q_1$ and $q_2$ in Eq.\ (\ref{eq:drive}) to obtain the ratchet effect in harmonic mixing and gating. 
Symmetry (\ref{eq:fr1}) holds for rocking ratchets induced by an additive bi-harmonic force, whereas 
(\ref{eq:fr2})  characterizes the gating average velocity. 
When $q_1+q_2$ is an odd integer number, the bi-harmonic force $f(t)$ breaks the time-shift symmetry 
$f(t)=-f(t+T/2)$ and a current appears. In a gating ratchet, if $q_1$ is an odd integer number, $f_1(t)$ preserves the time-shift symmetry, where $f_1(t)=-f_1(t+T/2)$. Nevertheless, the gating effect appears due to a 
 synchronization of the 
oscillations of the potential barrier caused by a single harmonic $f_1(t)$ with the 
motion produced by the additive harmonic force, $f_2(t)$. There is no constraint  on $q_2$ in gating, and therefore a  current can be obtained even for $q_1=q_2=1$ \cite{gommers:2008,zamora-sillero:2006}.

Moreover, Eqs.\ (\ref{eq:timeshift})-(\ref{eq:fr2}) together with the functional representation of the ratchet velocity determine  
the dependence of the current on the amplitudes and relative phase of the harmonics \cite{quintero:2010,cuesta:2013}. 
For instance, 
 for the small-amplitude limit of the bi-harmonic force $f(t)=f_{1}(t)+f_{2}(t)$ with (\ref{eq:drive}),  
 the current reads 
\begin{equation}\label{eq:velo}
v[f(t)]= A_{0} \epsilon_{1}^{q_{2}} \epsilon_{2}^{q_{1}} \cos(q_{1} \phi+\theta_{0}),
\end{equation}
where  $q_1+q_2$ is an odd integer number. Otherwise the current $v$ vanishes \cite{cuesta:2013,quintero:2010}.  The constants $A_{0}$  and $\theta_{0}$ are determined by the 
other parameters of the system (potential, dissipation, etc). Equation (\ref{eq:velo}) clearly shows the harmonic mixing since the parameters of the fist harmonic always appear in combination with the parameters of the second harmonic. 
Interestingly, for a gating ratchet, it is deduced (for a small-amplitude limit) that $v$ again is ruled by Eq.\ (\ref{eq:velo}), 
however only $q_1$ should be an odd integer number, whereas $q_2$ can be either an odd or even integer number. 
This formula predicts a sinusoidal dependence of $v$ versus 
the phase $\phi$. This implies, for example, that current reversals can be induced  by solely changing the relative phase 
between $f_1$ and $f_2$. 
Furthermore, in  \cite{cuesta:2013}, for a non-small amplitude limit, two interesting effects have been 
theoretically predicted:  
a  deviation from the sinusoidal shape of 
$v$ as a function of the phase; and the dependence of  
$\theta_{0}$ and $A_{0}$ on the amplitudes of the forces. 
This latter fact leads to an unexpected phenomenon related with the appearance of  \textit{current reversals  by changing the amplitudes of the harmonics}. This explains the experiments 
in optical lattices driven by a bi-harmonic force reported in \cite{cubero:2010}, and in a shaken liquid drop driven by two independent harmonics \cite{noblin:2009}.

In this work, we focus on the ratchet dynamics of  Brownian particles lying in 
a symmetric potential, modulated 
by a harmonic function. The particles are driven by an external 
  sinusoidal ({\em harmonic}) force.  We show that there is a deviation from the sinusoidal behavior of $v$ 
  as a function of the relative phase between the two harmonics in the non-small amplitude limit. 
  Moreover, the current reversals by means of increasing the amplitudes of the harmonics are shown.      

The paper is organized as follows: In the next Section, 
 the symmetry properties of the Langevin equation and its relation with the functional representation of the 
current predicted in \cite{cuesta:2013} are described. In Section III, 
  the analytical predictions of the previous section  are verified  
by means of simulations. In addition to a class of current reversals, 
determined by dissipation-induced symmetry breaking \cite{cubero:2010}, we 
show that the current reversals persist even for the overdamped dynamics of our model. 
To conclude the paper, in the last Section,   
the results of Sections II--III are discussed, thereby making the connection with the
experiments and summarizing our main findings.
 
\section{Gating ratchet model} \label{model}

In our theoretical analysis, the dynamics of particles in the  
spatially symmetric potential is determined by the Langevin equation
\begin{equation}\label{eq:langevin}
m\ddot{x} = -\alpha\dot{x}-U'(x) [1+f_{1}(t)] +f_{2}(t)+\sqrt{2 D} \xi (t),
\end{equation}
where $m$ is the mass; $U(x)= U_{0} \cos(x)$ is a periodic symmetric potential, modulated by the harmonic $f_{1}(t)$ given by (\ref{eq:drive}); 
$\alpha$ the friction
coefficient; $\xi(t)$ a Gaussian white noise, 
$\langle \xi(t)\rangle =0$, $\langle \xi(t)\xi(t')\rangle = \delta(t-t')$. Generally, noise smooths the dependence of the current on the parameters of the harmonics \cite{morales-molina:2005}. In some cases, as we show below, adding noise promotes transport. The additive force $f_{2}(t)$ are given by  
(\ref{eq:drive}).  All these magnitudes and 
parameters are in dimensionless form.

The current defined by Eq.\ (\ref{eq:v}) 
is time-shift invariant, i.e. it fulfils the symmetry (\ref{eq:timeshift}) due to the dissipation. Therefore, if 
$v$   is a {\it smooth} functional  
such that its functional Taylor series exists, then Theorem 1 of \cite{cuesta:2013} assures that 
\begin{equation}
v=\sum_{k=0}^\infty
(\epsilon_1^{q_2} \epsilon_2^{q_1})^kC_k(\epsilon_1,\epsilon_2) 
\cos\big( k q_1 \phi+\theta_k(\epsilon_1,\epsilon_2)\big),
\label{eq:vf}
\end{equation}
with $\theta_0(\epsilon_1,\epsilon_2)=0$, 
and functions
$C_{k}(\epsilon_1,\epsilon_2)$ and the phase lags 
$\theta_{k}(\epsilon_1,\epsilon_2)$ are even in each $\epsilon_j$, $j=1,2$. 
Notice that the symmetry (\ref{eq:fr2}) holds since exchanging $f_{2}$ with $-f_{2}$ 
is equivalent to replacing   $x(t)$ with $-x(t)$ in (\ref{eq:langevin}). The statistical properties of the Gaussian white noise are the same under the inversion of $\xi(t)$ to $-\xi(t)$. Therefore, all $C_k$ with 
 even $k$ are zero. With this restriction, the first two terms in (\ref{eq:vf}), for $q_1=q_2=1$,   read
 \begin{equation}\label{eq:11}
v=v_{1} \cos(\phi+\theta_{1})+ v_{2} \cos(3 \phi+\theta_{2})+E_{10}(\epsilon_{1},\epsilon_2),
 \end{equation}
 where $v_{1}=\epsilon_1 \epsilon_2 C_{1}(\epsilon_{1},\epsilon_2)$, 
  $v_2=(\epsilon_1 \epsilon_2)^3 C_{2}(\epsilon_{1},\epsilon_2)$, 
$C_{1}$ and $\theta_{1}$ are polynomials up to order 6 in $\epsilon_j$, and  $C_{2}$ and $\theta_2$ are linear in $\epsilon_1^2$ 
 and $\epsilon_2^2$. 
 
 For $q_1=1$ and $q_2=2$, $v$ is given by 
 \begin{equation} \label{eq:12}
v=v_{1} \cos(\phi+\theta_{1})+ v_{2} \cos(3 \phi+\theta_{2})+E_{15}(\epsilon_{1},\epsilon_2),
 \end{equation}
 where $E_{15}$ contains terms of order $15$ or higher in each $\epsilon_j$; 
 $v_{1}=\epsilon_1^2 \epsilon_2 C_{1}(\epsilon_{1},\epsilon_2)$, 
  $v_2=(\epsilon_1^2 \epsilon_2)^3 C_{2}(\epsilon_{1},\epsilon_2)$, $C_{1}$ and $\theta_{1}$ are even polynomials in $\epsilon_1$  and $\epsilon_2$ up to order $10$, and $C_{2}$ and $\theta_2$, are even polynomials in $\epsilon_1$  and $\epsilon_2$ up to order $4$. 
  In both cases,  ($q_1=q_2=1$ or $q_1=1$ and $q_2=2$) we have identified 3 main regimes which depend on the amplitudes of the harmonics, namely:  
 \begin{enumerate}
 \item Small-amplitude regime. Only the first term in (\ref{eq:11}) and (\ref{eq:12}) dominates  and $C_{1}$ and $\theta_{1}$ do not depend on the amplitudes. By fixing all the parameters of the system, $v$ is a sinusoidal function on $\phi$. 
 \item Intermediate amplitude regime. The second term in (\ref{eq:11}) and (\ref{eq:12}) can be neglected. However, in contrast 
 to the previous case, 
$C_{1}$ and $\theta_{1}$ do depend on $\epsilon_1$ and $\epsilon_2$. 
Therefore, the current reversals can be achieved by modifying the amplitudes. The sinusoidal behavior of $v$ persists. 
\item Large amplitude regime. The effect of the second term in (\ref{eq:11}) and (\ref{eq:12}) is observed, and therefore 
$v$ is no longer a sinusoidal function. 
 \end{enumerate}

The previous analysis remains valid for the overdamped dynamics. 
To describe the overdamped system we set $m \to 0$ in Eq. (\ref{eq:langevin}): 
\begin{equation}\label{eq:o}
\alpha\dot{x}=-U'(x) [1+f_{1}(t)] +f_{2}(t)+\sqrt{2 D} \xi (t).
\end{equation}
Moreover,  
 time-reversal now implies that by changing $f_1(t) \to f_{1}(-t)$, 
$f_{2}(t) \to f_{2}(-t)$ and $x(t) \to \pi-x(-t)$, the Eq. (\ref{eq:o}) remains 
invariant and 
\begin{equation}\label{eq:tr}
v[f_1(-t),f_{2}(-t)]=v[f_{1}(t),f_{2}(t)].
\end{equation} 
This symmetry fixes all the phase lags in Eq. (\ref{eq:vf}) to zero. 
Therefore, all the phase lags in Eqs.\ (\ref{eq:11}) and (\ref{eq:12}) are also zero.
Nevertheless,  current reversals may still be observed by changing the amplitudes of the forces. 
For instance, in the intermediate regime, a variation  in the parameters around the values for which $v_{1}(\epsilon_1,\epsilon_2)=0$ 
in Eqs.\ (\ref{eq:11}) and (\ref{eq:12}), could make  $v$ change its sign.     

In the following section,  all these findings are verified by means of  simulations of Eqs. (\ref{eq:langevin}) and (\ref{eq:o}).

\section{Simulations of the Langevin equation} \label{numerics}

Simulations of the stochastic differential Eqs.\ (\ref{eq:langevin}) and (\ref{eq:o}) have been performed using the Heun method and the 
2nd-order weak predictor-corrector method \cite{kloeden:1995}.  
The final time of integration is $2000$, the time step is either $0.1$ or $0.01$, and results are averaged over $10000$ realizations unless specified otherwise in the figure caption.

\begin{figure}[h]   
\centering
\includegraphics[width=8cm]{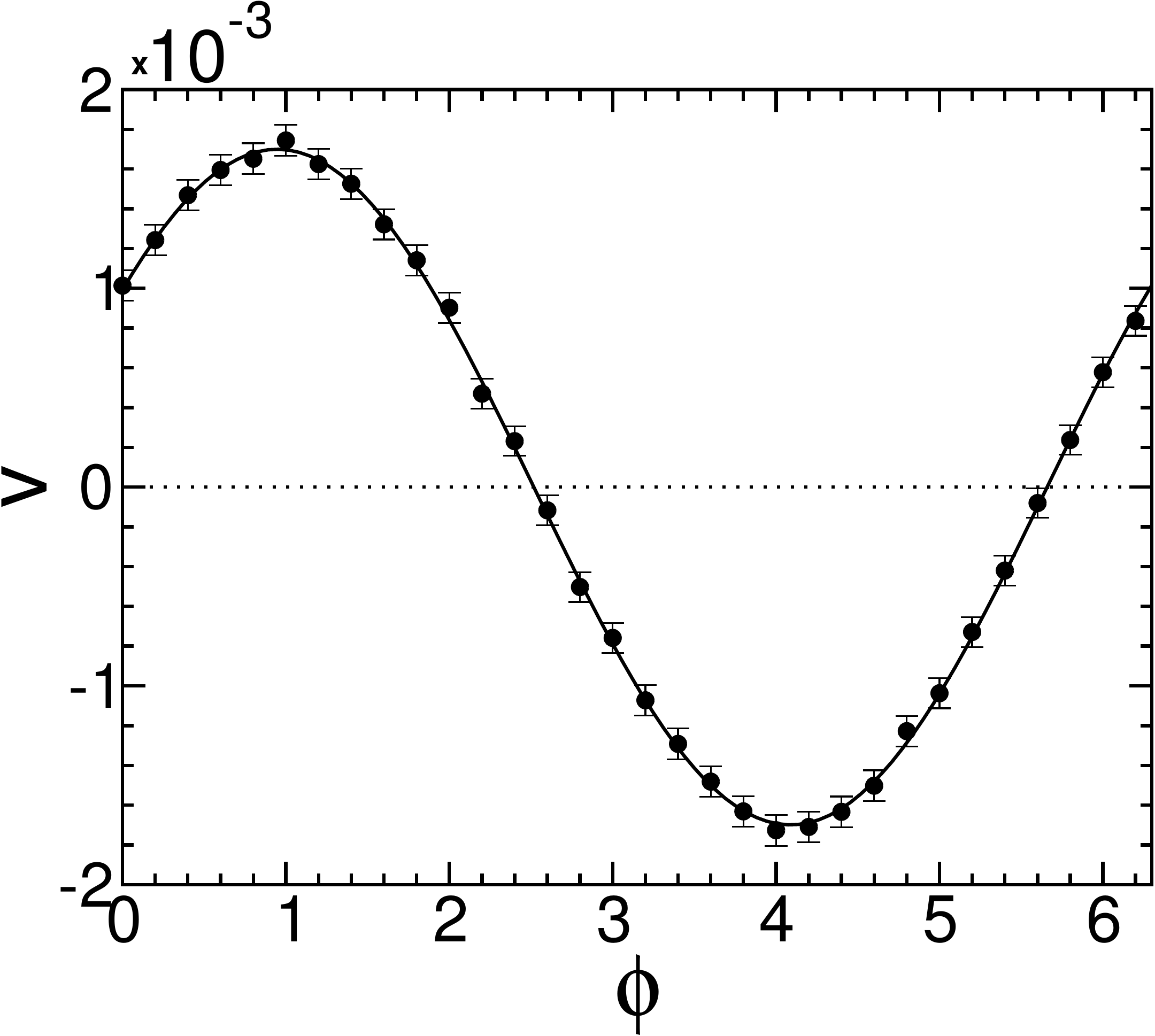}
\caption{$v$ vs $\phi$ from simulations of (\ref{eq:langevin}).  
Filled circles with error bars: current in steady state computed from the Eq.\ (\ref{eqv2}). 
Solid line represents the fitting curve of the circles, 
 $v=-0.00170 \cos(\phi+2.195)$.    
Parameters: $m=1$, $\alpha=1$, $U_0=5$, $\epsilon_1=\epsilon_2=0.5$, $\omega=1$,  $q_1=1$, $q_2=2$ and $D=1$.}
\label{fig2}
\end{figure}

The current $v$ is computed by means of  
\begin{equation}\label{eqv2}
 v=\left \langle \frac{x(t_f)-x(t_r)}{t_f-t_{r}} \right \rangle, 
\end{equation}
 where $t_r$ and $t_f$ are the final time of integration and the transient time, respectively 
(see Fig.\ \ref{fig2}).

\begin{figure}[h]
\centering
\includegraphics[width=8cm]{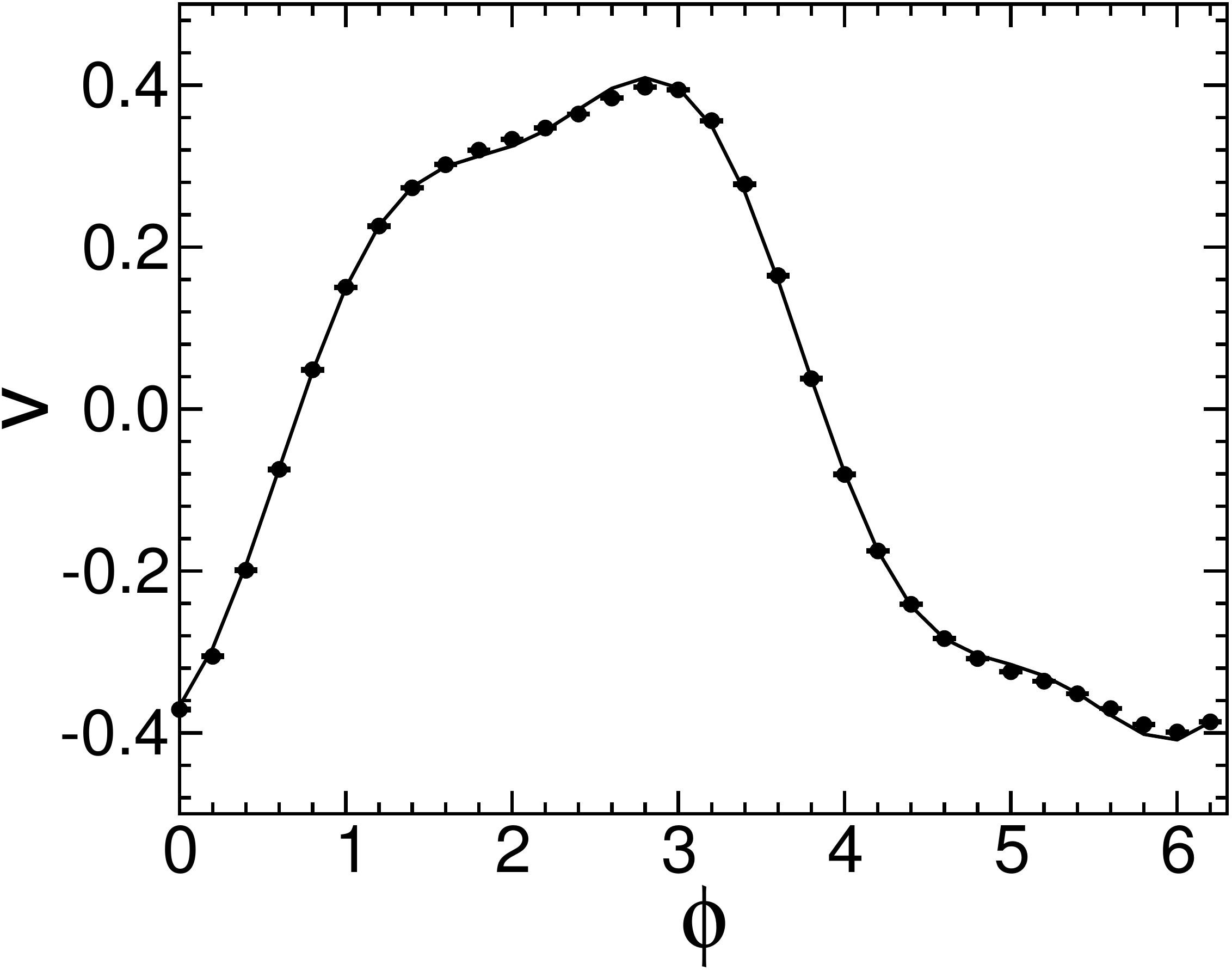}
\caption{$v$ vs $\phi$ from simulations of (\ref{eq:langevin}) shows non-sinusoidal behavior (filled circles with error bars).
Solid line represents the fitting curve $v=-0.4168 \cos(\phi+0.7713)-0.0686 \cos(3 \phi-0.0909)$. 
Parameters: $m=1$, $\alpha=1$, $U_0=5$, $\epsilon_1=\epsilon_2=2$, $\omega=1$,  $q_1=1$, $q_2=2$, and $D=1$. Final time of integration $1000$.}
\label{fig7}
\end{figure}

The  sinusoidal behavior of $v(\phi)$ is characteristic of the small and intermediate amplitude regimes. 
In Fig.\ \ref{fig2},  a sinusoidal behavior of $v$ is  
observed as a function of the phase. Close to $\phi \approx 2.5$ and $\phi \approx 5.7$, the velocity changes its sign and current reversals can appear by varying the phase and other parameters of the system that have an influence on the phase lag.

By  further increasing the amplitudes, the average velocity deviates from purely sinusoidal behavior and sinusoids of higher frequencies appear in its expansion. 
Indeed, in Fig.\ \ref{fig7}, the results from simulations of Eq.\ (\ref{eq:langevin}) can be fitted perfectly with two harmonics.
In Figs.\ (\ref{fig2}) and (\ref{fig7}), we notice that on replacing $\phi$ with $\phi+\pi$ (this is equivalent to replacing  
$f_2$ with $-f_2$), $v$ changes its sign. This means that the symmetry (\ref{eq:fr2}) is fulfilled.

\begin{figure}[h] 
\centering
\includegraphics[width=8cm]{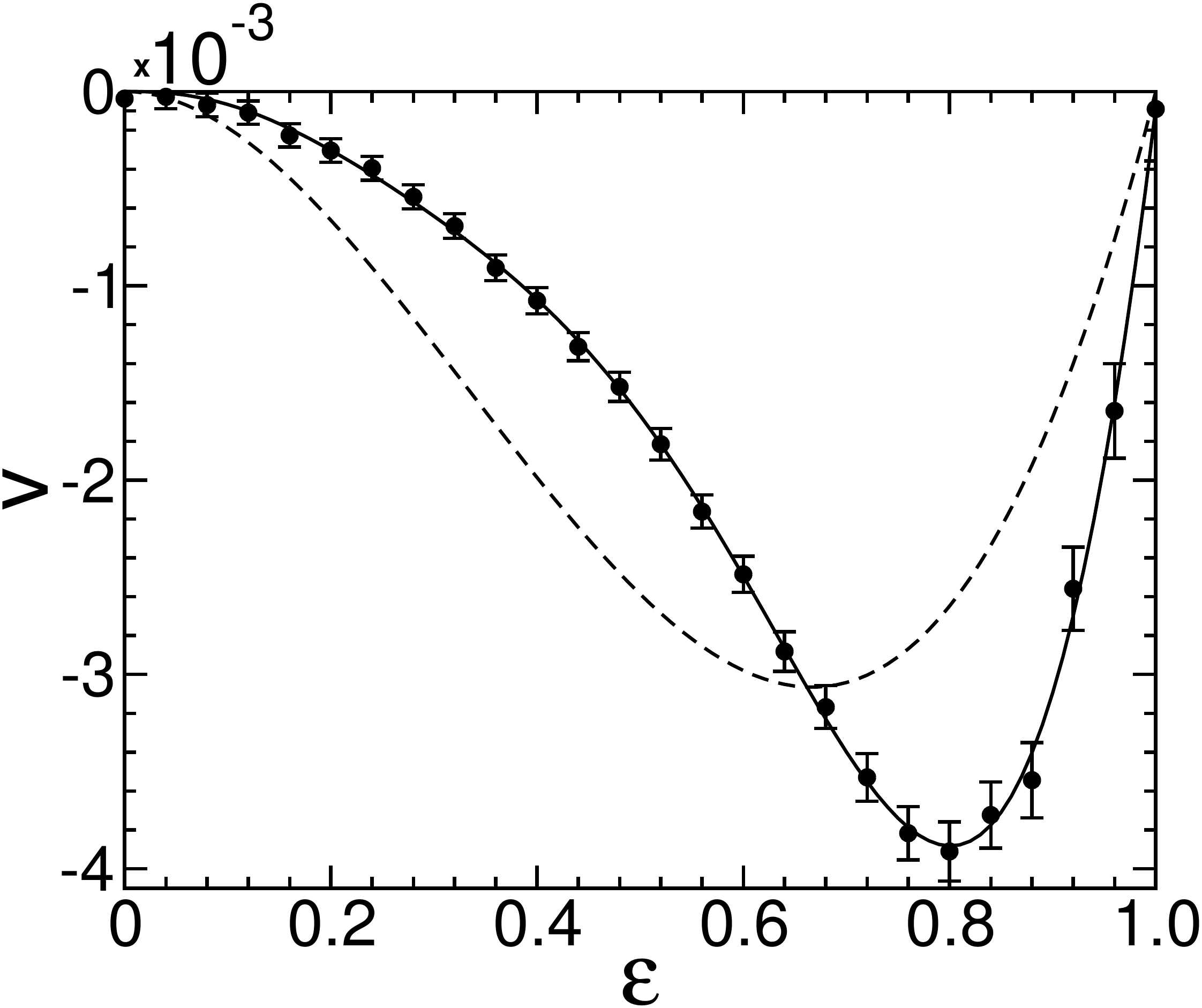}
\caption{$v$ vs $\epsilon$ from simulations of (\ref{eq:langevin}). 
Filled circles: current from the Eq.\ (\ref{eqv2}). 
Dashed line: fitting curve  $v=-0.0207 \epsilon^2 (1-\epsilon)$, 
 predicted for small-amplitude limit.   
 Solid line represents the fitting curve of the simulation points predicted in the intermediate amplitude regime, 
$v=\sum_{k=0}^{5} a_{k} \epsilon^{k+2}$, with $a_0 = -0.002$, $a_1 = -0.076$, $a_2 = 0.367$, $a_3=-0.747$, $a_4=0.664$, 
$a_5=-0.205$. 
Parameters: $m=1$, $\alpha=1$, $U_0=5$, $\epsilon_1=A \epsilon$, $\epsilon_2=A (1-\epsilon)$, $A=1$,  $\phi=4.09$, $\omega=1$, 
$q_1=1$, $q_2=2$, and $D=1$. }
\label{fig4}
\end{figure}

By fixing all the parameters, except $\epsilon_1$ and $\epsilon_2$ which vary according to $\epsilon_1=A \epsilon$, $\epsilon_2=A (1-\epsilon)$, we verify that the dependence of $v$ on $\epsilon$ is different from the expected $v \sim \epsilon^2 (1-\epsilon)$, which is valid for the small-amplitude regime  (see Fig. \ref{fig4}).
 
\begin{figure}[h] 
\centering
\includegraphics[width=8cm]{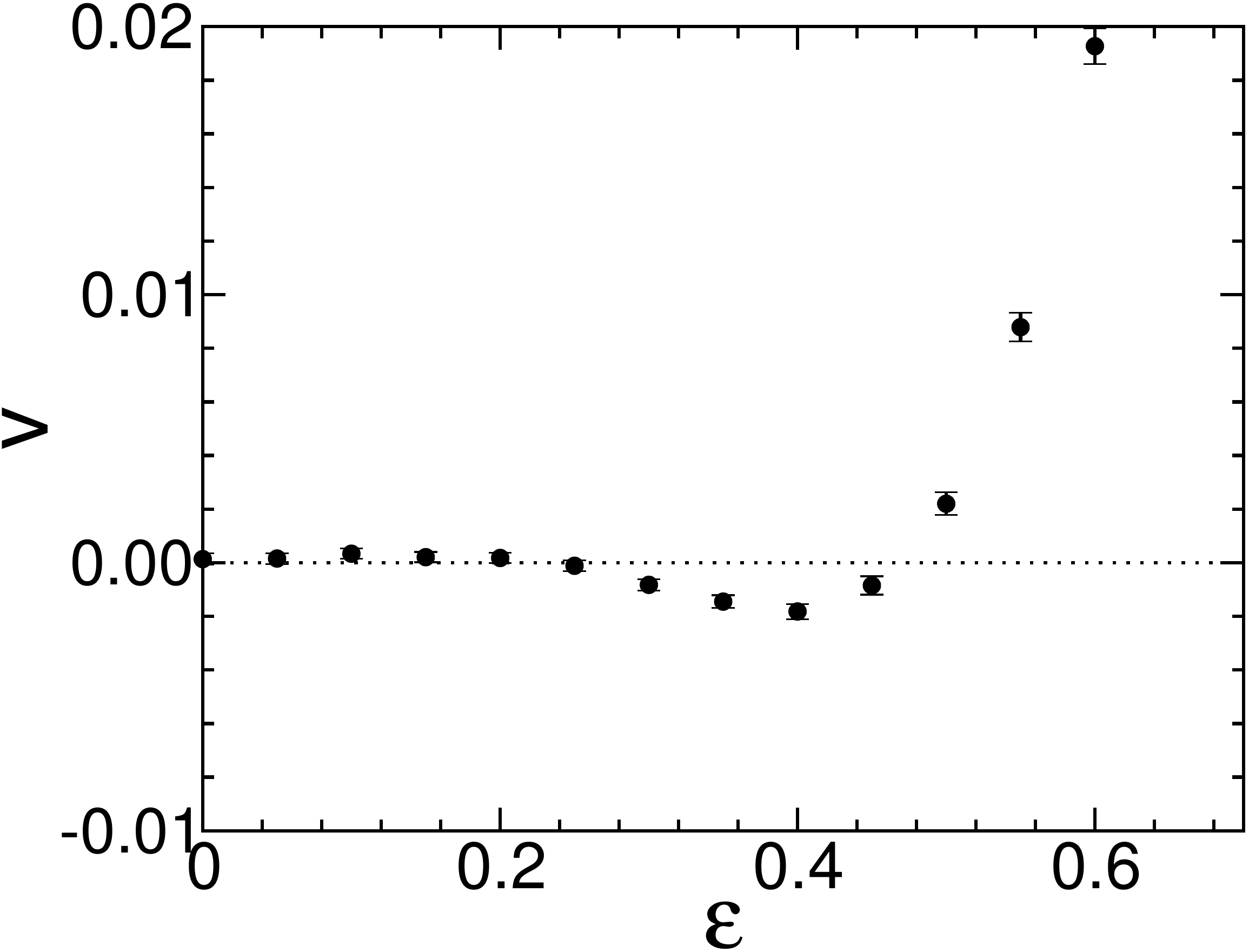}
\caption{$v$ vs $\epsilon$ from simulations of (\ref{eq:langevin}) shows that the direction of the current changes at approximately  
$\epsilon=0.5$.     
Parameters: $m=1$, $\alpha=1$, $U_0=5$, $\epsilon_1=2 \epsilon$, $\epsilon_2=2 (1-\epsilon)$, $\omega=1$, $\phi=2.8$, $q_1=1$, 
$q_2=2$, and $D=1$. Final time of integration $1000$. Dotted line represents zero velocity.}
\label{fig9}
\end{figure}

In order to observe a  current reversal via an amplitude change, 
first we fix all the parameters of Eq.\ (\ref{eq:langevin}) as in Fig.\ 
\ref{fig2}, except  
the amplitudes of the harmonics, which we have increased up to $\epsilon_1=\epsilon_2=1$.  The amplitudes are now sufficiently large for  the phase lags $\theta_k$  to be no longer constant and for them to depend on the amplitudes $\epsilon_1,\epsilon_2$ as in Eq. \eqref{eq:vf}. We set a relative phase $\phi\approx 2.8$ which corresponds to an almost vanishing current for $\epsilon_1=\epsilon_2=1$ (not shown in the figures). A clear current reversal appears by  modifying only the amplitudes around these values following  $\epsilon_1=2 \epsilon$, $\epsilon_2=2 (1-\epsilon)$ with $\epsilon\in[0,1]$, as shown in Fig.\ \ref{fig9}. The inversion of the current occurs around $\epsilon=0.5$, which corresponds to values $\epsilon_1=\epsilon_2=1$ and a vanishing $v$, as expected. 

\subsection{Overdamped dynamics of Brownian particle}

\begin{figure}[h] 
\centering
\includegraphics[width=8cm]{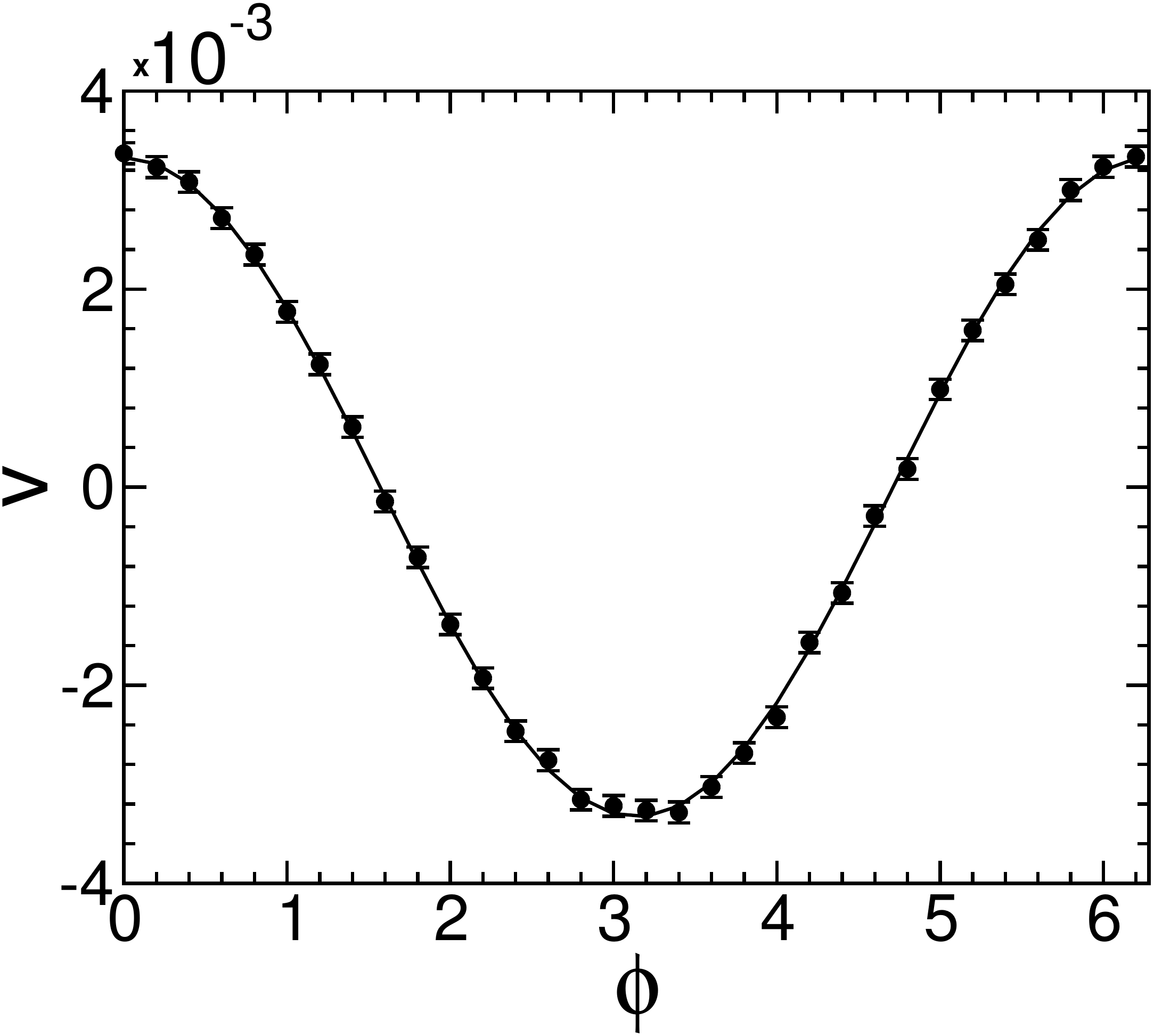}
\caption{$v$ vs $\phi$ from simulations of the overdamped system (\ref{eq:o}) (filled circles with error bars). 
Solid line represents the fitting curve  $v=0.0033 \cos(\phi)$.    
Parameters: $\alpha=1$, $U_0=5$, $\epsilon_1=\epsilon_2=0.5$, $\omega=1$,  $q_1=1$, $q_2=2$, and $D=1$. }
\label{fig3}
\end{figure}

 Interestingly, the maximum current shown in Fig.\ \ref{fig3} for the overdamped particle is greater 
than the maximum current reached when the inertial term remains in the Langevin equation, see Fig.\ 
\ref{fig2}. Notice that the parameters in both figures are the same, except the 
inertial term which is omitted in the simulations reported in Fig.\ \ref{fig3}. 
This effect resembles the enhancement of the movement due to the dissipation 
studied in \cite{salerno:2002,quintero:2010} in the relativistic particle driven by 
a bi-harmonic force.   

\begin{figure}[h] 
\centering
\includegraphics[width=8cm]{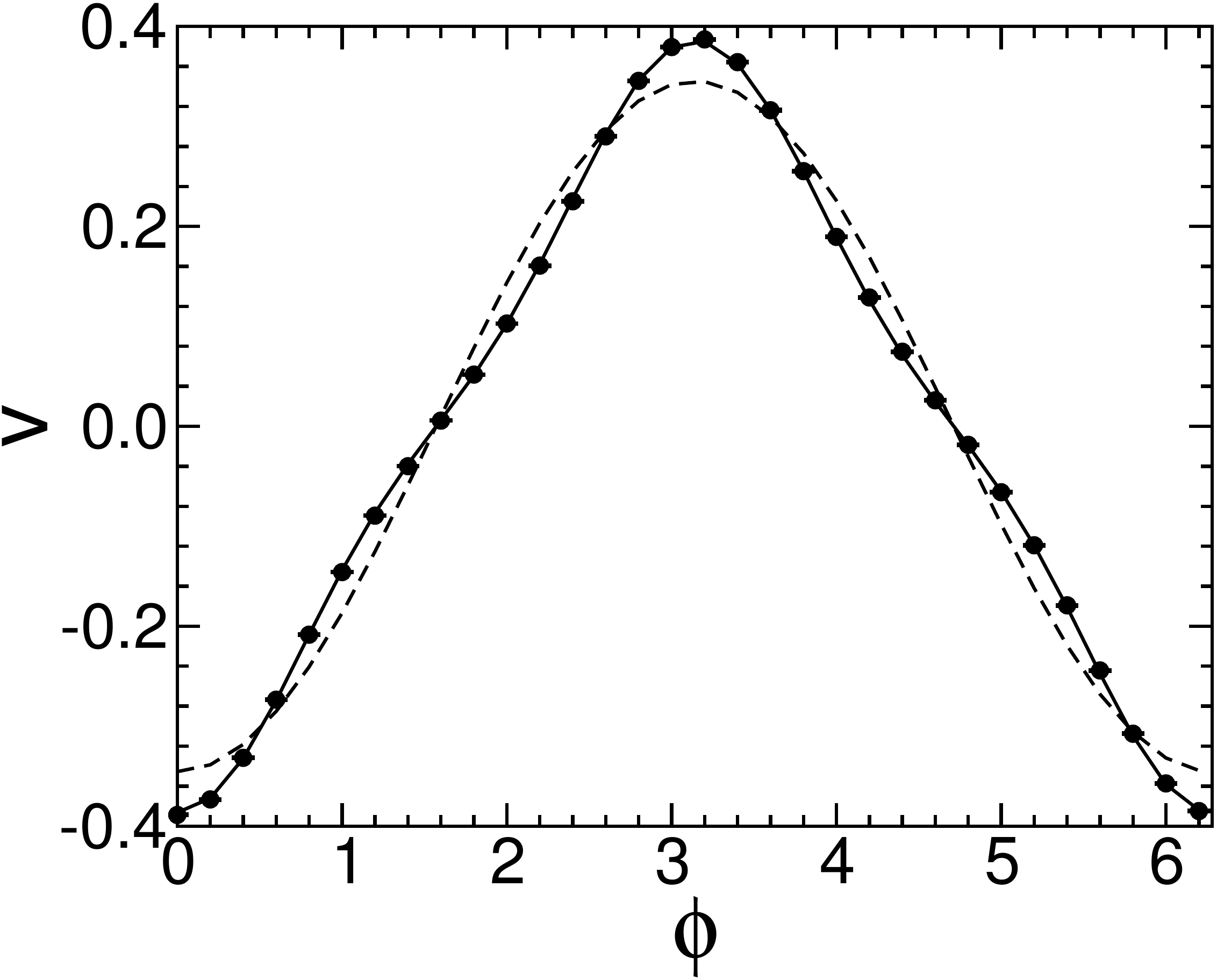}
\caption{$v$ vs $\phi$ from simulations of the overdamped system (\ref{eq:o}), 
filled circles with error bars.  
Dashed and solid lines are the fitting curves  
$v=-0.345 \cos(\phi)$ and $v=-0.344 \cos(\phi) -0.042 \cos(3 \phi)$, respectively.   
Parameters: $\alpha=1$, $U_0=5$, $\epsilon_1=\epsilon_2=2$, $\omega=1$, $q_1=1$, $q_2=2$, and $D=1$.}
\label{fig8}
\end{figure}

This striking phenomenon vanishes when the amplitudes are increased (the maxima  
of the currents shown in Figs.\  \ref{fig7} and \ref{fig8} are almost  
the same). On increasing the amplitudes, a small deviation from the sinusoidal behavior of $v$ 
as a function of the phase $\phi$ is also observed in the overdamped system, see Fig.\ \ref{fig8}.

\begin{figure}[h] 
\centering
\includegraphics[width=8cm]{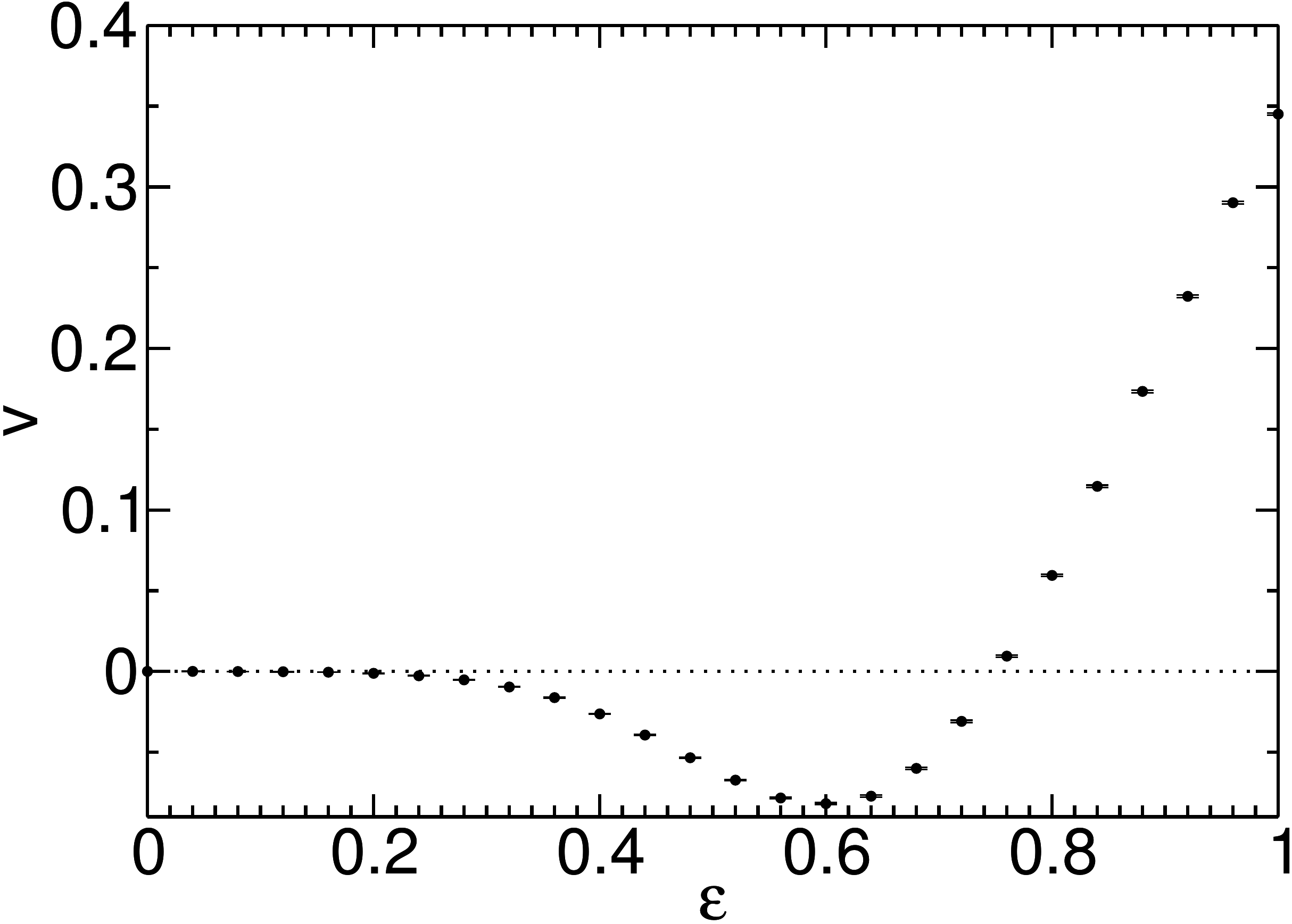}
\caption{$v$ vs $\epsilon$ from simulations of (\ref{eq:o}) shows that the direction of the current changes around 
$\epsilon \approx 0.75$. Amplitudes of $f_1$ and $f_2$ are varied simultaneously as  $\epsilon_1=2 \epsilon$, $\epsilon_2=2 \epsilon$. The rest of parameters are: $\alpha=1$, $U_0=5$, $\omega=1$, $q_1=1$, $q_2=2$, $\phi=2.8$, and $D=1$. $20000$ realizations.}
\label{fig11}
\end{figure}

\begin{figure}[h] 
\centering
\includegraphics[width=8cm]{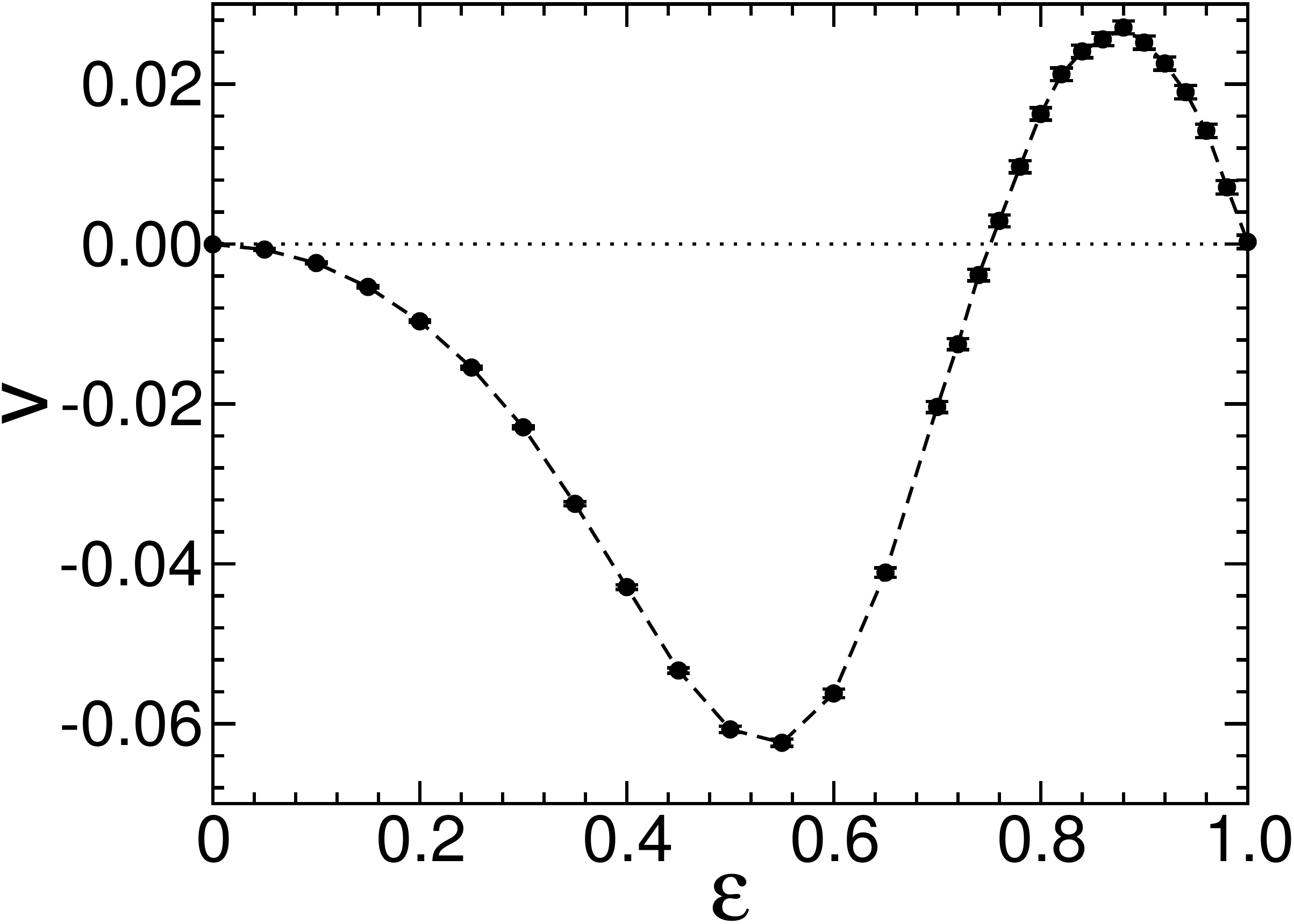}
\caption{$v$ vs $\epsilon$ from simulations of (\ref{eq:o}) shows that the direction of the current also changes around $\epsilon \approx 0.75$ (filled circles) when the amplitudes are varied with constant total amplitude $2$ as $\epsilon_1=2 \epsilon$, $\epsilon_2=2 (1-\epsilon)$.  The rest of parameters are $\alpha=1$, $U_0=5$, $\omega=1$, $q_1=1$, $q_2=2$, $\phi=2.8$, and $D=1$. $20000$ realizations.}
\label{fig_inve1e2_diff}
\end{figure}

In the overdamped dynamics, the phase lags are fixed to zero and the
search for current reversals associated to changes in the amplitudes of the harmonics   
becomes a more difficult task (see Fig.\ \ref{fig3}). 
In order to observe  a current reversal by changing the amplitudes of the forces,  
we must proceed in a different fashion. 
Results from simulations shown in Figs.\ 
\ref{fig3} and \ref{fig8} reveal  that, by changing the amplitudes $\epsilon_1$ and $\epsilon_2$ from 0.5 to 2 when $\phi \approx 3$, the direction of motion can be inverted. Therefore, by setting the phase,  for instance at $\phi=2.8$, and varying the amplitudes in the form of $\epsilon_1= 2\epsilon$, $\epsilon_2=\epsilon_1$ with $\epsilon \in [0,1]$, an inversion of the current is expected for a value of $\epsilon_1$ between $0.5$ and $2$. These results are shown in Fig. ~\ref{fig11}. Finally, Fig. ~\ref{fig_inve1e2_diff} shows that an inversion is also observed when the amplitudes are modified while keeping the total amplitude $\epsilon_1+\epsilon_2$ constant. Moreover, Fig.\ ~\ref{fig_inve1e2_diff} shows that $v=0$ when $\epsilon=0$ (no modulation of the potential) or  $\epsilon=1$ (no additive force).

\begin{figure}[h] 
\centering
\begin{tabular}{cc}
\includegraphics[width=3in]{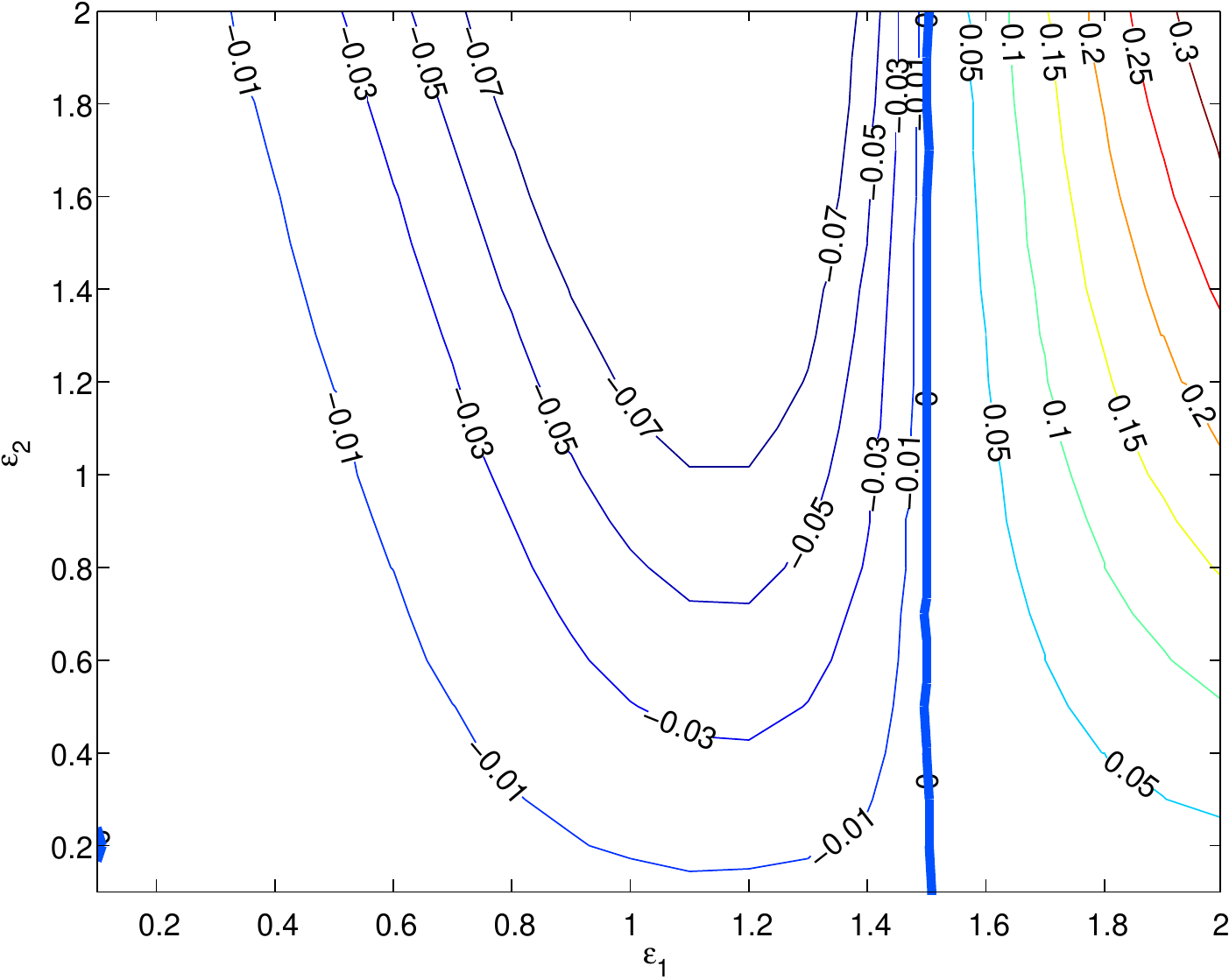} & \quad 
\includegraphics[width=3in]{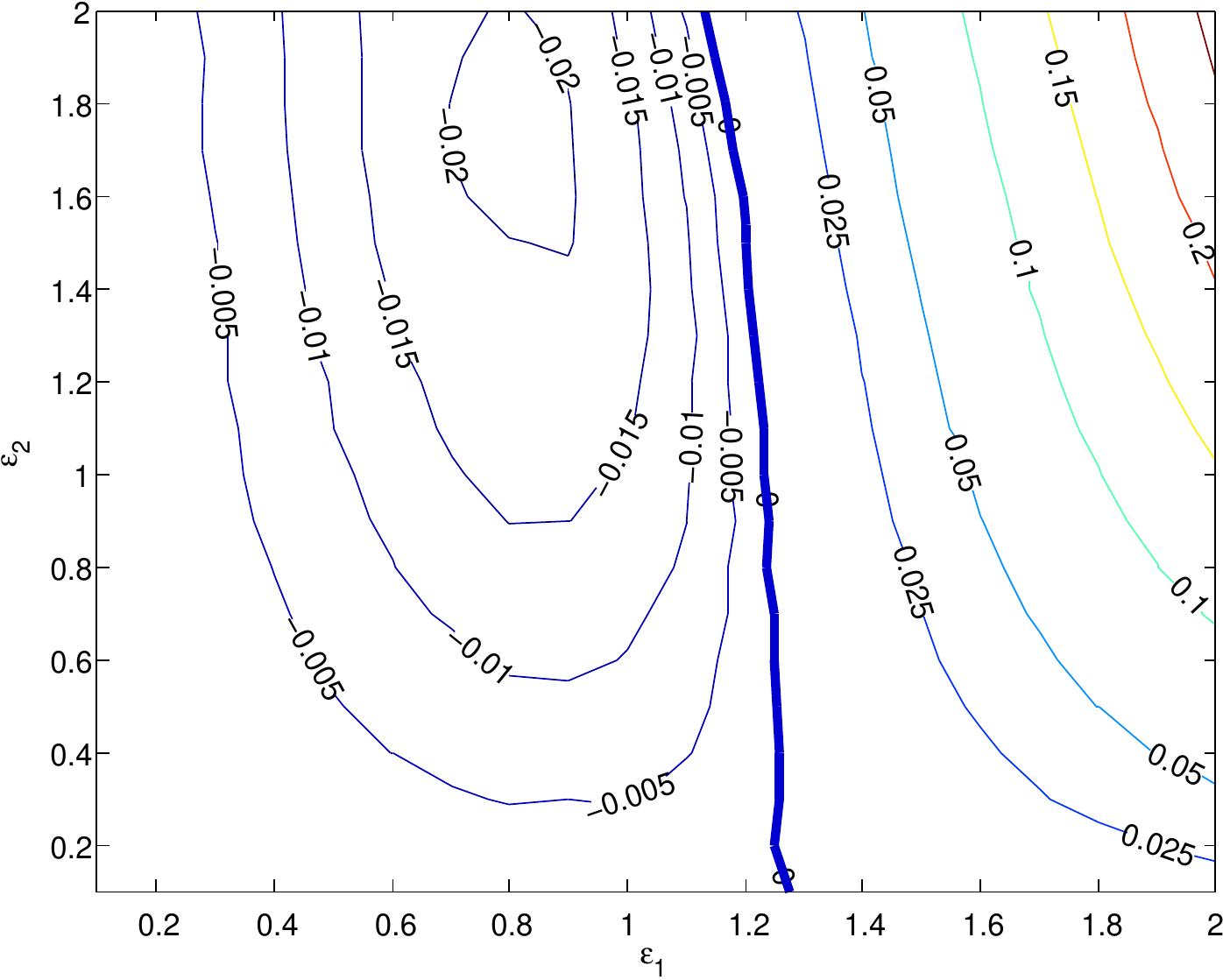}
\end{tabular} 
\caption{Contour plots $v$ as functions of $\epsilon_1$ and $\epsilon_2$ from simulations of (\ref{eq:o}). The thick
blue line marks the inversion of the current. Left panel $U_0=5$. Right panel: $U_0=2.5$. The rest of parameters are $\alpha=1$, $\omega=1$, $q_1=1$, $q_2=2$, $\phi=2.8$, and $D=1$.}
\label{new9}
\end{figure}

In Figs.\ \ref{fig11} and \ref{fig_inve1e2_diff}, the inversion of the current occurs at $\epsilon_1 \approx 1.5$. Indeed, by fixing all the parameters and changing $\epsilon_1$ and $\epsilon_2$,  
the contour plot (left panel in Fig. \ref{new9}) shows that the current vanishes when $\epsilon_1 \approx 1.5$. However, for other sets of parameters, for instance taking $U_0=2.5$ (see right-hand panel of Fig. \ref{new9}), the reversal current 
appears for different values of $\epsilon_1$.


\section{Summary} \label{conclu}

In this work, we study the 
dynamics of particles, driven by a harmonic force and subjected to white noise, when they are in a spatially symmetric potential that is modulated by a harmonic function. Both the applied force and the modulation of the potential  are time symmetric 
and the current $v$ fulfils the symmetry (\ref{eq:fr2}). Dissipation is also included in the description; therefore the current is time-shift invariant and the theory developed in 
 \cite{cuesta:2013} can be applied.

 We show that this theory predicts  three different regimes for our system which depend on the amplitudes of the two harmonics,  namely:  
 i) A small-amplitude regime where the current, 
 $v \sim \epsilon_1^{q_2} 
 \epsilon_2^{q_1} \cos(\phi+\theta_1)$, is a sinusoidal function with a phase lag, 
 $\theta_1$, independent of the amplitudes. This regime has been 
 predicted by the collective coordinate theory and confirmed by simulations in 
 the framework of soliton ratchets (see 
 \cite{zamora-sillero:2006} and references therein). ii) The intermediate 
 amplitude regime, where $v$ is still a sinusoidal function, although $\theta_1$ 
 is no longer a constant. This means that the sinusoidal behavior alone cannot  
 guarantee that the amplitudes of the harmonics are small. Therefore, in addition to experiments 
 on optical lattices reported in \cite{gommers:2008}, in order to determine 
 the regime where the system lies, it is necessary to investigate the dependence of $v$ on the amplitudes 
 $\epsilon_1$ and $\epsilon_2$. Once the intermediate regime 
 is reached, current reversals via an amplitude change is expected. This phenomenon is confirmed 
 by simulations of the underdamped Langevin Eq.\ (\ref{eq:langevin}). It is worthy of note that 
 current reversals have been found to be present in the overdamped limit, where the current satisfies the time-reversal symmetry (\ref{eq:tr}).  
iii) Large-amplitude regime, where we show that the non-sinusoidal 
behavior of the current, predicted by the theory, is due to the increasing strength of the two harmonics. 
 
Apart from the results presented in Figs.1--8, we have also performed
simulations for all the set of parameters of Figs.\ \ref{fig2}-\ref{fig_inve1e2_diff}, but we fixed  
the strength of the noise $D=0$. In all cases, the computed current is zero (of order of $10^{-9}$ or less). 
Therefore, for the set of parameters studied here the noise  together with the action of the harmonics generate  the transport.   
 
Finally, it is pointed out that, according to the theory developed in \cite{cuesta:2013}, the main phenomena studied here using a specific model, can appear in other physical systems that satisfy the same symmetries, 
 including experimental realizations 
 in Josephson junctions \cite{falo:2002,ustinov:2004,beck:2005} 
 and optical lattices \cite{schiavoni:2003}, in which a number of the above results have been reported. Other results, however, 
   require verification through experiments.

\section*{Acknowledgments}
We acknowledge financial support through: grants 
FIS2011-24540 (N.R.Q.) and ENFASIS (L.D.); from
Ministerio de Econom\'{\i}a y Competitividad (Spain); grants 
FQM207 (N.R.Q.), and P09-FQM-4643 (N.R.Q.), from Junta de
Andaluc\'{\i}a (Spain); and especially a grant from the Alexander von Humboldt Foundation (Germany) through Research
Fellowship for Experienced Researchers SPA 1146358 STP (N.R.Q.). Part of the calculations of this work were performed in the high
capacity cluster for physics, funded in part by UCM and  in part with 
Feder FUNDS. This is a contribution to the Campus of International 
Excellence of Moncloa, CEI Moncloa. 

\bibliographystyle{plain}
\bibliography{ratchets}

\end{document}